\documentclass[conference, letterpaper]{IEEEtran2014}

\usepackage[latin1]{inputenc}
\usepackage{graphicx}
\usepackage{amsthm}
\usepackage{amssymb}
\usepackage{amsmath}
\usepackage{cite}
\usepackage{url}
\usepackage{balance}
\usepackage{array}
\usepackage{booktabs}

\usepackage{algorithm}
\usepackage{algpseudocode}

\usepackage{mathtools}
\DeclarePairedDelimiter{\ceil}{\lceil}{\rceil}
\DeclarePairedDelimiter{\floor}{\lfloor}{\rfloor}

\usepackage{epstopdf}

\newtheorem{mytheorem}{Theorem}
\IEEEoverridecommandlockouts

\begin{document}

\title{Effect of Wideband Beam Squint on Codebook Design in Phased-Array Wireless Systems}

\author{\IEEEauthorblockN{
Mingming Cai\IEEEauthorrefmark{1},
Kang Gao\IEEEauthorrefmark{1},
Ding Nie\IEEEauthorrefmark{1},
Bertrand Hochwald\IEEEauthorrefmark{1},
J. Nicholas Laneman\IEEEauthorrefmark{1},\\
Huang Huang\IEEEauthorrefmark{2} and
Kunpeng Liu\IEEEauthorrefmark{2}}
\IEEEauthorblockA{\IEEEauthorrefmark{1}Wireless Institute, University of Notre Dame\\
Email: \texttt{\{mcai, kgao, nding1, bhochwald, jnl\}@nd.edu}}
\IEEEauthorblockA{\IEEEauthorrefmark{2}Huawei Technologies Co., Ltd.\\
Email: \texttt{\{huanghuang, liukunpeng\}@huawei.com}}
\thanks{This work has been supported by Huawei Technologies Co., Ltd.}
}

\maketitle

\begin{abstract}
Analog beamforming with phased arrays is a promising technique for 5G wireless communication at millimeter wave frequencies. Using a discrete codebook consisting of multiple analog beams, each beam focuses on a certain range of angles of arrival or departure and corresponds to a set of fixed phase shifts across frequency due to practical hardware considerations. However, for sufficiently large bandwidth, the gain provided by the phased array is actually frequency dependent, which is an effect called beam squint, and this effect occurs even if the radiation pattern of the antenna elements is frequency independent.  This paper examines the nature of beam squint for a uniform linear array (ULA) and analyzes its impact on codebook design as a function of the number of antennas and system bandwidth normalized by the carrier frequency.  The criterion for codebook design is to guarantee that each beam's minimum gain for a range of angles and for all frequencies in the wideband system exceeds a target threshold, for example 3~dB below the array's maximum gain.  Analysis and numerical examples suggest that a denser codebook is required to compensate for beam squint.  For example, 54\% more beams are needed compared to a codebook design that ignores beam squint for a ULA with 32 antennas operating at a carrier frequency of 73~GHz and bandwidth of 2.5~GHz. Furthermore, beam squint with this design criterion limits the bandwidth or the number of antennas of the array if the other one is fixed.
\end{abstract}

\section{Introduction}
\label{sec:introduction}
%

Two major directions are being explored to enhance radio frequency (RF) spectrum access, namely, reusing currently under-utilized frequency bands below roughly 6~GHz through dynamic spectrum access (DSA) and exploiting spectrum opportunities in higher frequencies, e.g., millimeter-wave (mmWave) bands \cite{mingmingcaiJNLAsilomar2015, el2014spatially}. The total amount of spectrum that can be shared below 6~GHz is limited to hundreds of MHz, whereas mmWave bands can provide spectrum opportunities on the order of tens of GHz.  However, the signals at mmWave frequencies experience significantly higher path loss than signals below 6~GHz \cite{rappaporttheodores2002}.

Beamforming with a phased array having a large number of antennas is a promising approach to compensate for higher attenuation at mmWave frequencies while supporting mobile applications \cite{huang2015joint,hur2013millimeter, heath2015overview}.
First, the short wavelength of mmWave allows for integration of a large number of antennas into a phased array of small size suitable for a mobile wireless device.
Second, switched beamforming applies different sets of parameters to the phased array, allowing the beam direction to be electronically and quickly controlled, which enables steering for different orientations and tracking for mobility.  In this paper, we consider switched analog beamforming with one RF chain using a phased array with phase shifters.  For hybrid architectures, phased arrays can also be implemented by either phase shifters or switches \cite{mendez2015hybrid}.

Traditionally in a communication system with analog beamforming, the sets of phase shifter values in a phased array are designed at a specific frequency, usually the carrier frequency, but applied to all frequencies within the transmission bandwidth due to practical hardware constraint. Phase shifters are relatively good approximations to the ideal time shifters for narrowband transmission; however, this approximation breaks down for wideband transmission if the angle of arrival (AoA) or angle of departure (AoD) is away from broadside, because the required phase shifts are frequency-dependent.
The net result is that beams for frequencies other than the carrier ``squint'' as a function of frequency \cite{mailloux2005phased} in a wide signal bandwidth. This phenomenon is called beam squint in a wideband system \cite{SeyedGarakkoui2011BeamSquinting}. As we will see in the context of orthogonal division multiplexing (OFDM) modulation with multiple subcarriers, beam squint translates into array gains for a given AoA or AoD that vary with subcarrier index.  To eliminate or reduce beam squint with hardware approaches, true time delay and phase improvement schemes have been proposed  \cite{longbrake2012true, liu2013minimize}; however, these approaches are unappealing for mobile wireless communication due to combinations of high implementation cost, significant insertion loss, large size, or excessive power consumption.

Covering a certain range of AoA or AoD in switched beamforming requires a ``codebook'' consisting of multiple beams, with each beam determined by a set of beamforming phases \cite{song2015codebook, cui2014evolution}. Usually, a minimum array gain is required for the codebook to cover the entire desired angle range \cite{hehierarchical}, for example, 3~dB below the array's maximum gain. A natural extension for a wideband system is to design the beamforming codebook to guarantee the minimum gain for all frequencies, even in the presence of beam squint.  To the best of our knowledge, only \cite{liu2013minimize} proposes a phase improvement scheme to reduce the effect of beam squint in beamforming codebook, but with high implementation cost of additional phase shifters and bandpass filters in mmWave frequency.
By contrast, we propose developing denser codebooks to compensate for beam squint at the system level.

The main contributions of this paper are twofold.  First, we model the beam squint for a uniform linear array (ULA) in a wireless communication system, and show that the usable beamwidth for each beam effectively decreases because of the beam squint. Therefore, the number of beams in the codebook must increase to counter this effect.
Second, we develop a codebook design algorithm to enforce minimum array gain for all subcarriers in the wideband system. The bandwidth of the system is limited for a fixed number of antennas in the phased array; conversely, the number of antennas in the phased array is limited for a fixed bandwidth. The algorithm and conclusions also apply to frequencies other than mmWave and acoustic communication.

\section{System Model}
\begin{figure}
\begin{center}
\includegraphics[width=85 mm]{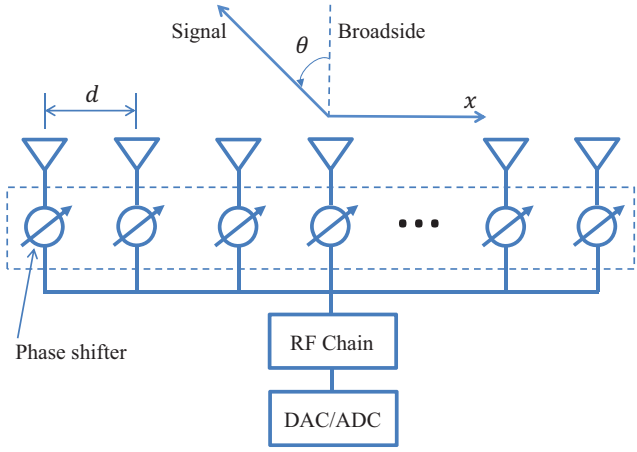}
\end{center}
\caption{Structure of a uniform linear array (ULA) with analog beamforming using phase shifters.  The distance between the adjacent antennas is $d$, and $\theta$ denotes either the angle-of-arrival (AoA) for reception or the angle-of-departure (AoD) for transmission.}
\label{fig:array-structure}
\end{figure}

We consider a ULA as shown in Fig.~\ref{fig:array-structure}. It could be the phased array of either a transmitter or receiver. In the array, there are $N$ identical and isotropic antennas, each connected to a phase shifter. The distances between two adjacent antenna elements are the same and are all denoted by $d$. For simplicity, we also assume the phases of the phase shifters are continuous without quantization. The AoD or AoA $\theta$ is the angle of the signal relative to the broadside of the array, increasing counterclockwise.
From \cite{orfanidis2002electromagnetic}, the ULA response vector is
\begin{align}
{\mathbf{a}}\left( \theta \right) =
\left[ 1,{e^{j2\pi {\lambda ^{ - 1}}d\sin \theta}}, \ldots ,{e^{j2\pi {\lambda ^{ - 1}}nd\sin\theta }}, \ldots , \right. \nonumber \\
 \left. {e^{j2\pi {\lambda ^{ - 1}}\left( {N - 1} \right)d\sin \theta}} \right]^T, \ \theta \in {\left[ -\frac{\pi}{2}, \frac{\pi}{2} \right]},
\label{eq:receive-array-response-linear-array}
\end{align}
where $\lambda$ is the signal wavelength, and superscript $\left( \cdot \right) ^T$ indicates transpose of a vector. The effect of the phase shifters can be modeled by the beamforming vector
\begin{align}
{\mathbf{w}} = \left[ {e^{j\beta _1},\dots,e^{j\beta_n},\dots,e^{j\beta_N}} \right]^T,
\label{eq:beam-forming-vector}
\end{align}
where $\beta_n$ is the phase of $n$th phase shifter connected to the $n$th antenna element in the array.

From \cite{balanis2016antenna}, the array factors or array gains for both the transmitter and receiver have the same expression
\begin{align}
g \left({\mathbf{w}},\theta \right)=\frac{1}{\sqrt {N}} {\mathbf{w}}^H  {\mathbf{a}}\left( \theta \right) = \frac{1}{\sqrt {N}} \sum\limits_{n = 1}^{N} {{e^{j\left[ {2\pi {\lambda ^{ - 1}}\left( {n - 1} \right)d\sin \theta - {\beta _n}} \right]}}}.
\label{eq:array-gain}
\end{align}
where superscript $\left( \cdot \right) ^H$ denotes Hermitian transpose.

\section{Modeling of Beam Squint in a Phased Array}

The analysis of the effect of beam squint is the same for a transmit or a receive array, so without loss of generality, we focus on the case of a receive array. Define the \emph{beam focus angle} $\theta_0\left(\mathbf{w}\right)$ to be the AoA for which the beam has the highest gain for a given beamforming vector $\mathbf{w}$. The beamforming vector $\mathbf{w}$ is traditionally designed for the carrier frequency. From (\ref{eq:array-gain}), to achieve the highest array gain for a beam focus angle $\theta_0$ among all possible $\mathbf{w}$, the phase shifters should be
\begin{align}
{\beta _n} \left( \theta_0 \right)=  2\pi {\lambda_c^{ - 1}}\left( {n - 1} \right)d\sin \theta_0, \ n=1,2,\dots,N,
\label{eq:phases-required-for-fine-beams}
\end{align}
where $\lambda_c$ is the wavelength of the carrier frequency. We call the beam with phase shifts described in (\ref{eq:phases-required-for-fine-beams}) a \emph{fine beam}, and we focus on the analysis of fine beams throughout the paper.

Based on (\ref{eq:array-gain}), the array gain for a signal with frequency $f$ and AoA $\theta$ using beam focus angle $\theta_0$ is
\begin{align}
g \left({\theta_0,  \theta, f}\right) = \frac{1}{{\sqrt {{N}} }}   \sum\limits_{n = 1}^N {{e^{j\left[ {2\pi {f c^{-1}}\left( {n - 1} \right)d\sin {\theta} - {\beta _n}\left(\theta_0 \right)} \right]}}},
\label{eq:ms-phase-combiner-implementation-freq}
\end{align}
where $c$ is the speed of light.

\begin{figure}
\begin{center}
\includegraphics[width=88 mm]{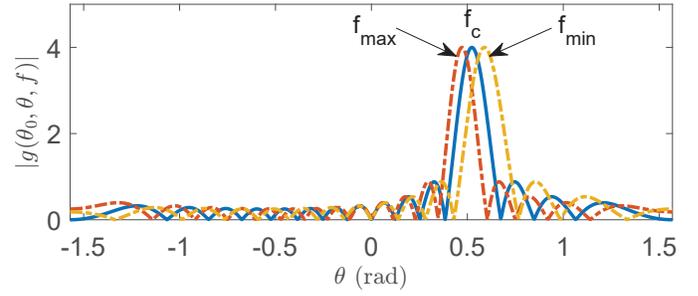}
\end{center}
\caption{Example of beam patterns varying for different frequencies in a wideband system with beam focus angle $\theta_0=\pi/6$, $N=16$ antennas, antenna spacing $d=\lambda_c/2$, carrier frequency$f_c=73$~GHz, minimum frequency $f_{min}=65.7$~GHz, and maximum frequency $f_{max}=80.2$~GHz. The bandwidth is chosen to be quite large for purposes of illustration.}
\label{fig:beam-plot-several-frequencies}
\end{figure}

For a fixed array and beamforming vector, an example of the beam squint patterns for different frequencies is shown in Fig.~\ref{fig:beam-plot-several-frequencies}. The beam focus angle for the carrier frequency is $\pi/6$. The array gain for an AoA $\theta$ are different for different frequencies; in particular, the array gains of $f_{max}$ and $f_{min}$ for the beam focus angle $\theta_0$ are both smaller than that of $f_{c}$.

For a wideband system, we express a subcarrier's absolute frequency $f$ relative to the carrier frequency $f_c$ as
\begin{align}
f=\xi f_c,
\label{eq:freq-coefficient}
\end{align}
where $\xi$ is the ratio of the subcarrier frequency to the carrier frequency. Suppose the baseband bandwidth of the signal is $B$. Then $f \in \left[ f_c-\frac{B}{2}, f_c+\frac{B}{2} \right]$. Define the fractional bandwidth
\begin{align}
b:={B}/{f_c},
\label{eq:fractional-bandwidth}
\end{align}
with $b>0$, so that $\xi \in \left[1-b/2, 1+b/2\right]$. As a specific example, $\xi$ varies from 0.983 to 1.017 in a system with 2.5~GHz bandwidth at 73~GHz carrier frequency.

The expression in (\ref{eq:ms-phase-combiner-implementation-freq}) can be transformed  into
\begin{align}
g \left({\theta_0,  \theta_c, \xi}\right) &= \frac{1}{{\sqrt {{N}} }}  \sum\limits_{n = 1}^{N} {{e^{j\left[ {2\pi {\xi f_c c^{-1}}\left( {n - 1} \right)d\sin {\theta_c}- {\beta _n}\left(\theta_0 \right)} \right]}}} & \nonumber\\
&=\frac{1}{{\sqrt {{N}} }}  \sum\limits_{n = 1}^{N} {{e^{j\left[ {2\pi { \lambda_c ^{-1}}\left( {n - 1} \right)d \left(\xi \sin {\theta_c} \right) - {\beta _n}\left(\theta_0 \right)} \right]}}},
\label{eq:ms-phase-combiner-implementation-ratio}
\end{align}
where $\theta_c$ is AoA for the carrier frequency.

Let ${\theta^{\prime}}$ be the equivalent AoA for the subcarrier with coefficient $\xi$. Comparing (\ref{eq:ms-phase-combiner-implementation-ratio}) to (\ref{eq:ms-phase-combiner-implementation-freq}), $\theta^{\prime}$ can be expressed as
\begin{align}
\theta^{\prime} = \arcsin \left( {\xi \sin \theta_c} \right).
\end{align}
The variation of beam patterns for different frequencies in the wideband system can therefore be interpreted as the variation of AoAs in the same angle response for different subcarriers.

\begin{figure}

\begin{center}
\includegraphics[width=88 mm]{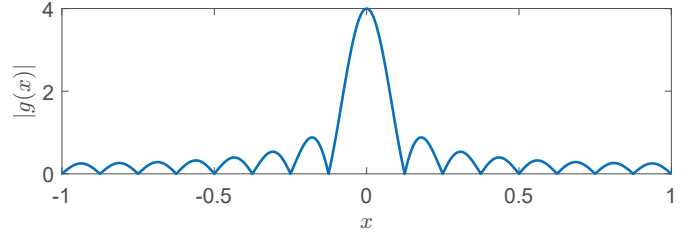}
\end{center}
\caption{Example of $\left|g \left(x \right) \right|$ from (\ref{eq:ms-array-gain-one-parameter}) for a fine beam with $N=16$ antennas and antenna spacing $d=\lambda_c/2$.}
\label{fig:Beam_shape_plot_fine_beam_psi}
\end{figure}

Usually, the beamforming array is designed such that $d=\lambda_c/2$. From (\ref{eq:phases-required-for-fine-beams}) and (\ref{eq:ms-phase-combiner-implementation-ratio}),
\begin{align}
g & \left({\theta_0,  \theta_c, \xi}\right) = \frac{1}{{\sqrt {{N}} }}  \sum\limits_{n = 1}^{N} {{e^{j\left[ {\pi \left( {n - 1} \right) \left(\xi \sin {\theta_c}-\sin {\theta_0} \right) } \right]}}} \nonumber \\
&= \frac{{\sin \left( {\frac{{N\pi }}{2}\left( {\xi \sin \theta_c  - \sin \theta_0 } \right)} \right)}}{{\sqrt N \sin \left( {\frac{\pi }{2}\left( {\xi \sin \theta_c  - \sin \theta_0 } \right)} \right)}} e^{j \frac{\left(N-1\right) \pi} {2} \left( {\xi \sin \theta_c- \sin \theta_0} \right)}.
\label{eq:ms-phase-combiner-implementation-ratio-3}
\end{align}
Define
\begin{align}
g \left(x \right) = \frac{{\sin \left( {\frac{{N\pi x}}{2}} \right)}}{{\sqrt N \sin \left( {\frac{\pi x }{2}} \right)}} e^{j \frac{\left(N-1\right) \pi x} {2}}.
\label{eq:ms-array-gain-one-parameter}
\end{align}
The array gain for frequency with $\xi$ in (\ref{eq:ms-phase-combiner-implementation-ratio-3}) becomes $g \left(\xi \sin \theta_c  - \sin \theta_0 \right)$. Fig.~\ref{fig:Beam_shape_plot_fine_beam_psi} illustrates an example plot of $\left|g \left(x \right) \right|$. The main lobe is the lobe with the highest gain.

Let
\begin{align}
\psi=\sin \theta
\end{align}
to convert the angle $\theta$ space to the $\psi$ space. Correspondingly, $\psi_c=\sin \theta_c$, $\psi_0=\sin \theta_0$, and $\psi_0 \left(\mathbf{w}\right)=\sin{\theta_0 \left(\mathbf{w}\right)}$. The array gain of a fine beam in (\ref{eq:ms-phase-combiner-implementation-ratio-3}) then becomes
\begin{align}
g \left(\xi \psi_c - \psi_0\right) =\frac{{\sin \left( {\frac{{N\pi }}{2}\left( {\xi  \psi_c  -  \psi_0 } \right)} \right)}}{{\sqrt N \sin \left( {\frac{\pi }{2}\left( {\xi \psi_c  - \psi_0 } \right)} \right)}} e^{j \frac{\left(N-1\right) \pi} {2} \left( {\xi \psi_c-  \psi_0} \right)},
\label{eq:ms-phase-combiner-implementation-ratio-4}
\end{align}
where $\varphi_c, \varphi_0 \in \left[-1,1\right]$. The maximum array gain is
\begin{align}
g_m=\mathop {\max }\limits_{\psi_c \in \left[-1,1\right]}   \left| g \left({\psi_c-\psi_0}\right) \right|=\sqrt{N}.
\label{maximum-array-gain}
\end{align}

For simplicity, we study the effect of beam squint in $\psi$ space. From (\ref{eq:ms-phase-combiner-implementation-ratio-4}), the effect of beam squint increases if the AoA is away from the beam focus angle, and it tends to increase as the beam focus angle increases.


\section{Codebook Design}
One widely used codebook design criterion is to require the array gain to be larger than a threshold $g_t$ for all subcarriers \cite{hehierarchical}. The threshold may vary for different system designs. In communication systems, the beam focus angle ${\psi _0}\left(\mathbf{w}\right)$ is determined by the beamforming vector $\mathbf{w}$. Define the beam coverage of $\mathbf{w}$ as the set
\begin{align}
\mathcal{R}\left( \mathbf{w} \right):
=&\left\{ \psi_c \in \left[ -\psi_m, \psi_m \right]: \right. \nonumber \\
&\left.\ \min_{\xi \in \left[1-b/2, 1+b/2\right] } \left| {g \left( { \xi \psi_c  -  {\psi _0}\left( {{{\mathbf{w}}}} \right) } \right)}\right| \ge {g_t} \right\},
\label{beam-range-definition}
\end{align}
where $0<\psi_m \le 1$ is the maximum targeted covered AoA/AoD of the codebook. The beam coverage of some beams is not contiguous in $\psi$ if $\psi_0$ is around -1 or 1. We only consider the beam whose beam range is contiguous.

Define the effective beamwidth $\Delta \psi$ as the measure of $\mathcal{R}\left(\mathbf{w}\right)$, i.e.,
\begin{align}
\Delta \psi  := \int_{ - {\psi _m}}^{{\psi _m}} {{I_{{\mathcal R}\left( {\mathbf{w}} \right)}}} \left( \psi  \right) d\psi,
\label{beam-width-definition}
\end{align}
where
\begin{align}
{I_{{\mathcal R}\left( {\mathbf{w}} \right)}}\left( \psi  \right) = \left\{
{\begin{array}{*{20}{l}}
   {1,\ \text{if}\ \psi  \in {\mathcal R}\left( {\mathbf{w}} \right)}  \\
   {0,\ \text{otherwise} }. \\
\end{array}} \right.
\end{align}

A single beam can only cover a small range of angles in space. In communication systems with beamforming, multiple beams with different beam focus angles are utilized to cover the target angle range $\left[ -\psi_m, \psi_m \right], 0<\psi_m \le 1$, forming a codebook $\mathcal{C}$. Suppose there are $M$ beams in the codebook to cover $\left[ -\psi_m, \psi_m \right], 0<\psi_m \le 1$. The codebook $\mathcal{C}$ has size $M=\left| \mathcal{C} \right|$ and can be defined as
\begin{align}
{\mathcal C} := \left\{ {{{\bf{w}}_m},m = 1,2,\dots,M:\bigcup\limits_{m = 1}^M {{\cal R}\left( {{{\mathbf{w}}_m}} \right)}  = \;\left[ { - {\psi _m},{\psi _m}} \right]} \right\}.
\label{codebook-definition}
\end{align}

Denote $\mathbf{\Omega}$ as the set of all codebooks that meet the requirements of (\ref{codebook-definition}). There is an infinite number of codebooks in $\mathbf{\Omega}$. Among them, we focus on the ones that achieve the minimum codebook size. Minimum codebook size is defined as
\begin{align}
{\left| \mathcal{C} \right| }_{min} =  \mathop {\min }\limits_{\mathcal{C} \in \mathbf{\Omega}} {{\left| \mathcal{C} \right| }}.
\end{align}

The threshold $g_t$ can be different for different system requirements. For illustration, we choose $g_t$ such that
\begin{align}
g_t={g_m}/ \sqrt{2},
\label{threshold-definition}
\end{align}
where $g_m$ is defined in (\ref{maximum-array-gain}). The analysis of other thresholds $g_t$ is similar.

\subsection{Codebook Design without Considering Beam Squint}
\begin{algorithm}[t]
\caption{Codebook Design without Beam Squint}
\label{codebook-design-no-beam-squint}
\begin{algorithmic}[1]
\State Calculate ${\left| \mathcal{C} \right| }_{min}$ from (\ref{minimum-codebook-size-no-beam-squint})
\If {${\left| \mathcal{C} \right| }_{min}$ is odd}
    \State Let $\psi_0=0$, $k=1$
    \State Align the first beam so that $\psi_0=0$
\Else
    \State Let $\psi_0= -\Delta\psi_{3\text{dB}}/2$, $k=0$
\EndIf
\State $\psi_0=\psi_0+\Delta\psi_{3\text{dB}}$
\While {$k < {\left| \mathcal{C} \right| }_{min}$}
    \State $k=k+1$
    \State Align the $k$th beam with focus angle $\psi_0$
    \State $k=k+1$
    \State Align the $k$th beam with focus angle $-\psi_0$
    \State $\psi_0=\psi_0+\Delta\psi_{3\text{dB}}$
\EndWhile
\end{algorithmic}
\end{algorithm}

Define beam edges as the points with $\psi$ in the main lobe whose array gain is $g_t$. There are two beam edges, one on the left side and one on the right side, denoted as $\psi_l$ and $\psi_r$, respectively. The selected $g_t$ in (\ref{threshold-definition}) is 3~dB below the maximum array gain $g_m$. The corresponding beamwidth is called the 3~dB beamwidth, denoted as $\Delta\psi_{3\text{dB}}$ in $\psi$ space. For consecutive beam coverage, the beamwidth is the distance between the two beam edges, i.e.,
\begin{align}
\Delta\psi_{3\text{dB}}=\psi_r-\psi_l.
\end{align}
According to \cite{orfanidis2002electromagnetic},
\begin{align}
\Delta\psi_{3\text{dB}}=\frac{1.772}{N},
\label{beam-width-N}
\end{align}
which is a constant independent of $\psi_0$. It is worth stressing that the beamwidth in $\theta$ space is not fixed, and depends on $\theta_0$.

To design a codebook with the minimum codebook size, we simply tile the beam coverages over $\left[ -\psi_m, \psi_m\right]$ without overlap.  If the effect of beam squint is not considered, all beams have the same beamwidth $\Delta\psi_{3\text{dB}}$ as shown in (\ref{beam-width-N}), and the minimum codebook size
\begin{align}
{\left| \mathcal{C} \right| }_{min}= \ceil[\bigg]{\frac{2\psi_m}{\Delta\psi_{3\text{dB}}}},
\label{minimum-codebook-size-no-beam-squint}
\end{align}
where $\ceil[\big]{\cdot}$ denotes the well-known ceiling function.

Even for the same ${\left| \mathcal{C} \right| }_{min}$, there could be multiple codebook designs. Among them, we are more interested in the ones that are symmetric relative to broadside. A codebook design method for minimum codebook size without considering the beam squint is shown in Algorithm~\ref{codebook-design-no-beam-squint}.


\subsection{Beam Pattern with Beam Squint}
As analyzed in the previous section, the array gains of some subcarriers are smaller than that of the carrier frequency when $\psi_c \ne 0$. To guarantee a minimum array gain $g_t$ for all the subcarriers, the beam range $\mathcal{R}\left( \mathbf{w} \right)$ is reduced.

Since $\mathbf{w}$ determines the beam focus angle $\psi_0$, $\mathcal{R}\left( \mathbf{w} \right)$ and $\mathcal{R}\left( \psi_0 \right)$ can be used interchangeably. Define
\begin{align}
&{\psi _{c,l}\left(\psi_0\right)}=\min \mathcal{R}\left( \psi_0 \right), \\
&{\psi _{c,r}\left(\psi_0\right)}=\max \mathcal{R}\left( \psi_0 \right).
\end{align}
First, we study the case for $\psi_c>0$, which gives us
\begin{align}
&{\psi _{c,l}\left(\psi_0\right)}\left( {1 - b/2} \right) = {\psi _l \left(\psi_0\right)}, \\
&{\psi _{c,r}\left(\psi_0\right)}\left( {1 + b/2} \right) = {\psi _r \left(\psi_0\right)},
\end{align}
where $\psi_l\left(\psi_0\right)$ and $\psi_r\left(\psi_0\right)$ indicate the left edge and the right edge of the beam with beam focus angle $\psi_0$ when there is no beam squint, i.e., $b=0$. Set $b=0$ in (\ref{beam-range-definition}), and we have
\begin{align}
{\psi _l \left(\psi_0\right)}=\psi_0- \frac{\Delta\psi_{3\text{dB}}}{2},\\
{\psi _r \left(\psi_0\right)}=\psi_0+ \frac{\Delta\psi_{3\text{dB}}}{2},
\end{align}
we have
\begin{align}
&{\psi _{c,l}\left(\psi_0\right)}= \frac{\psi_0- \frac{\Delta\psi_{3\text{dB}}}{2}}{ {1 - \frac{b}{2} } }, \label{psi-cl-calculation} \\
&{\psi _{c,r}\left(\psi_0\right)}= \frac{\psi_0+ \frac{\Delta\psi_{3\text{dB}}}{2}}{{1 +\frac{b}{2}  }  }, \label{psi-cr-calculation} \\
&\psi_0={\left( {1 -\frac{b}{2}} \right)} {\psi _{c,l}} + \frac{\Delta\psi_{3\text{dB}}}{2}.
\label{psi0-calculation}
\end{align}

Similarly, for $\psi_c<0$, we can obtain
\begin{align}
&{\psi _{c,l}\left(\psi_0\right)}= \frac{\psi_0- \frac{\Delta\psi_{3\text{dB}}}{2}}{ {1 +\frac{b}{2} } }, \label{psi-cl-calculation-psic-less-0} \\
&{\psi _{c,r}\left(\psi_0\right)}= \frac{\psi_0+ \frac{\Delta\psi_{3\text{dB}}}{2}}{{1 -\frac{b}{2}  }  }. \label{psi-cr-calculation-psic-less-0}
\end{align}

In considering effect of beam squint, the effective beamwidth for a beam with focus angle $\psi_0$ is
\begin{align}
\Delta\psi_{3\text{dB}} \left(\psi_0\right)&={\psi _{c,r}\left(\psi_0\right)}-{\psi _{c,l}\left(\psi_0\right)} \nonumber \\
&= \left\{
    {\begin{array}{{l}}
    {\frac{{\Delta {\psi _{3{\rm{dB}}}} - b{\left|\psi _0\right|}}}{{1 - \frac{{{b^2}}}{4}}}, \quad {\psi _l}\left( {{\psi _0}} \right){\psi _r}\left( {{\psi _0}} \right) > 0}  \\
    {\frac{{\Delta {\psi _{3{\rm{dB}}}}}}{{1 + \frac{b}{2}}},\quad \quad \quad \ {\psi _l}\left( {{\psi _0}} \right){\psi _r}\left( {{\psi _0}} \right) \le 0,}  \\
    \end{array}}
    \right.
\label{effective-beam-width}
\end{align}
where $\Delta\psi_{3\text{dB}}$ is calculated with (\ref{psi-cl-calculation}) and (\ref{psi-cr-calculation}), or (\ref{psi-cl-calculation-psic-less-0}) and (\ref{psi-cr-calculation-psic-less-0}) if ${\psi _l}\left( {{\psi _0}} \right),{\psi _r}\left( {{\psi _0}} \right) > 0$; otherwise, $\Delta\psi_{3\text{dB}}$ is calculated with (\ref{psi-cr-calculation}) and (\ref{psi-cl-calculation-psic-less-0}). If ${\psi _l}\left( {{\psi _0}} \right),{\psi _r}\left( {{\psi _0}} \right) > 0$, $\Delta\psi_{3\text{dB}} \left(\psi_0\right)$ is not fixed, and it is a linear function of $\psi_0$ for fixed $b$.


Since $\Delta\psi_{3\text{dB}} \left(\psi_0\right) >0$ for all $\psi_0 \in \left[ -\psi_m, \psi_m \right]$, based on (\ref{beam-width-N}) and (\ref{effective-beam-width}), we have the following theorem.
\begin{mytheorem}
If the number of antennas in the array $N$ is fixed, the fractional bandwidth $b$ is upper bounded by
\begin{equation}
b\le \frac{{\Delta\psi_{3\text{dB}}} }{\psi_m }=\frac{1.772}{\psi_m N};
\label{b-upper-bounded}
\end{equation}
Conversely, if the fractional bandwidth $b$ is fixed, the number of antennas in the array $N$ is upper bounded by
\begin{equation}
N \le \floor[\bigg]{\frac{1.772}{\psi_m b}},
\label{N-upper-bounded}
\end{equation}
where $\floor[\big]{\cdot}$ is a floor function.
\label{theorem:b-N-upper-bounded}
\end{mytheorem}


\subsection{Codebook Design with Beam Squint}


\begin{algorithm}[t]
\caption{Codebook Design with Beam Squint}
\label{codebook-design-beam-squint}
\begin{algorithmic}[1]
\Procedure{1}{Odd-Number Codebook}
\State Let $k=1$, $\psi_0=0$
\State Align the first beam so that $\psi_0=0$
\State Calculate $\psi _{c,r}$ based on (\ref{psi-cr-calculation})
\While {$\psi _{c,r}<\psi_m$}
    \State Let $\psi _{c,l}=\psi _{c,r}$
    \State Calculate $\psi_0$ based on (\ref{psi0-calculation})
    \State Calculate $\psi _{c,r}$ based on (\ref{psi-cr-calculation})
    \If {$\psi _{c,l} \ge \psi _{c,r}$}
        \State Codebook does not exist; algorithm ends
    \EndIf
    \State $k=k+1$
    \State Align the $k$th beam with focus angle $\psi_0$
    \State $k=k+1$
    \State Align the $k$th beam with focus angle $-\psi_0$
\EndWhile
\State $\left| \mathcal{C}_{o} \right|=k$
\EndProcedure

\Procedure{2}{Even-Number Codebook}
\State Let $k=0$, $\psi _{c,r}=0$
\While {$\psi _{c,r}<\psi_m$}
    \State Let $\psi _{c,l}=\psi _{c,r}$
    \State Calculate $\psi_0$ based on (\ref{psi0-calculation})
    \State Calculate $\psi _{c,r}$ based on (\ref{psi-cr-calculation})
    \State $k=k+1$
    \State Align the $k$th beam with focus angle $\psi_0$
    \State $k=k+1$
    \State Align the $k$th beam with focus angle $-\psi_0$
\EndWhile
\State $\left| \mathcal{C}_{e} \right|=k$
\EndProcedure
\State ${\left| \mathcal{C} \right| }_{min}=\min\left(\left| \mathcal{C}_{o} \right| , \left| \mathcal{C}_{e} \right| \right)$
\State Select the Procedure that achieves ${\left| \mathcal{C} \right| }_{min}$
\end{algorithmic}
\end{algorithm}

\begin{figure}
\begin{center}
\includegraphics[width=88 mm]{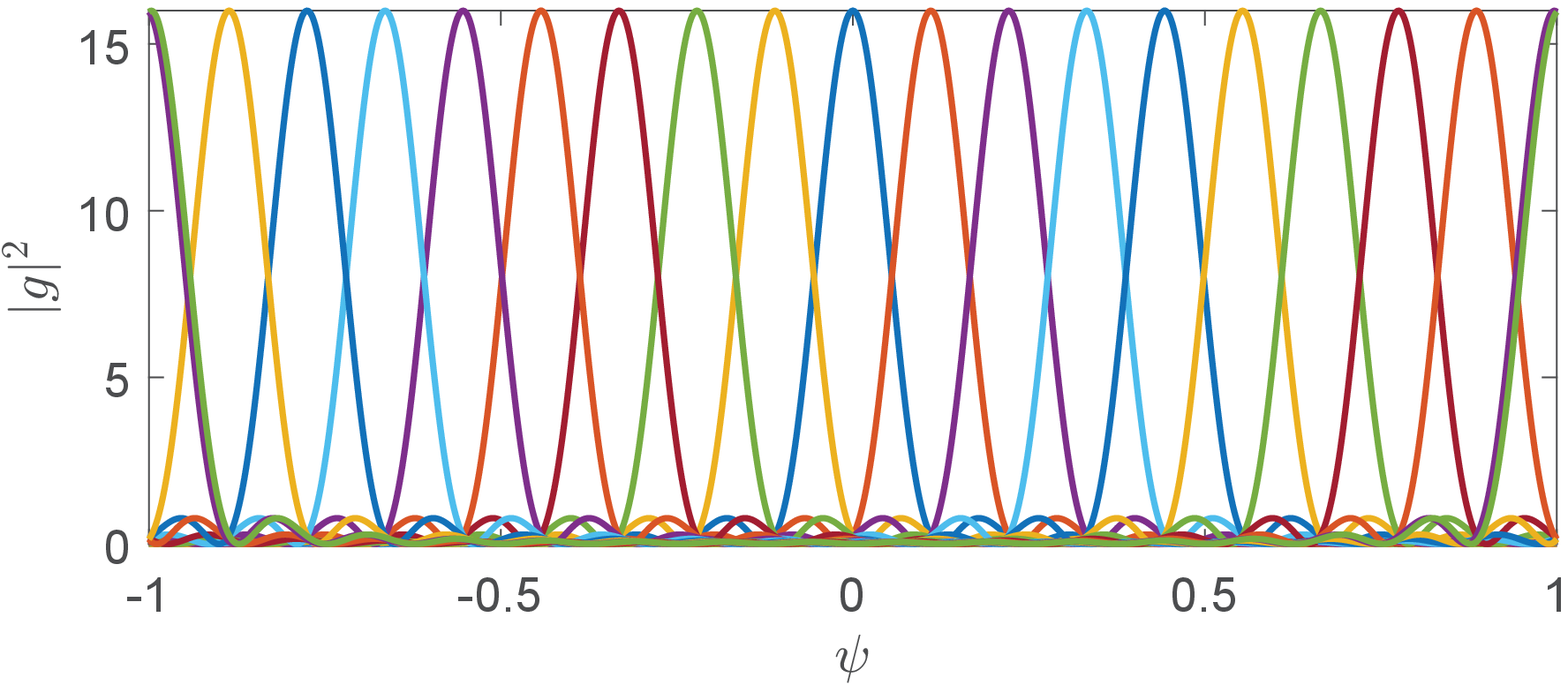}
\end{center}
\caption{Example of a codebook design for the minimum codebook size without consideration of beam squint according to Algorithm~\ref{codebook-design-no-beam-squint}. The two beams with the largest $\psi_0$ are located closely. ${\left| \mathcal{C} \right| }_{min}=19$. $N=16$, $\psi_m=1$, $d=\lambda_c/2$.}
\label{fig:codebook-design-without-wideband-effect}
\bigskip
\begin{center}
\includegraphics[width=88 mm]{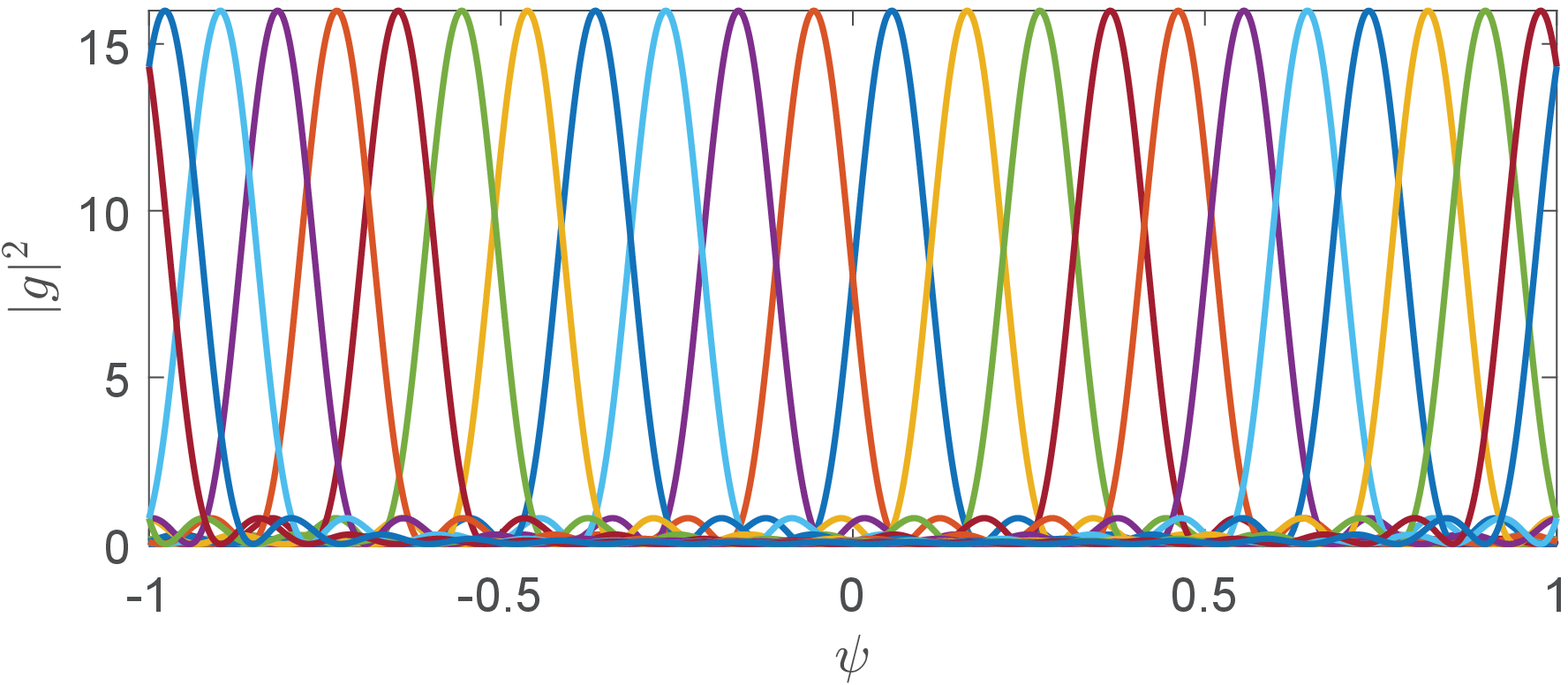}
\end{center}
\caption{Example of a codebook design for the minimum codebook size with consideration of beam squint according to the Algorithm~\ref{codebook-design-beam-squint}. The beams in the figure are for the carrier frequency. ${\left| \mathcal{C} \right| }_{min}=22$. $N=16$, $\psi_m=1$, $d=\lambda_c/2$. $b=0.0342$ corresponds to $B=$2.5 GHz at $f_c=$73 GHz.}
\label{fig:codebook-design-example}
\end{figure}

Again, to achieve the minimum codebook size, the beam coverages in the codebook should have as little overlap as possible. Algorithm~\ref{codebook-design-beam-squint} shows one method to design a codebook that achieves minimum codebook size with beams symmetric to broadside. Before running the algorithm, it is not clear whether the codebook contains an odd or an even number of beams that achieve the minimum codebook size. Procedure~1 in Algorithm~\ref{codebook-design-beam-squint} designs an odd-number codebook, denoted as ${\mathcal{C}_{o} }$; Procedure~2 designs an even-number codebook, denoted as ${\mathcal{C}_{e} }$. For aligning the beams on the right of the broadside, ${\psi _{c,r}\left(\psi_0\right)}$ of the beam with smaller beam focus angle is equal to ${\psi _{c,l}\left(\psi_0\right)}$ of the beam with larger beam focus angle. The first determined beam in Algorithm~\ref{codebook-design-beam-squint} has $\psi_0=0$.


Either ${\mathcal{C}_{o} }$ or ${\mathcal{C}_{e}}$ achieves the minimum codebook size. The minimum codebook size is also calculated in Algorithm~\ref{codebook-design-beam-squint}, i.e.,
\begin{align}
{\left| \mathcal{C} \right| }_{min}=\min\left(\left| \mathcal{C}_{o} \right| , \left| \mathcal{C}_{e} \right| \right).
\end{align}

\section{Numerical Result}

\begin{figure}[t]
\begin{center}
\includegraphics[width=88 mm]{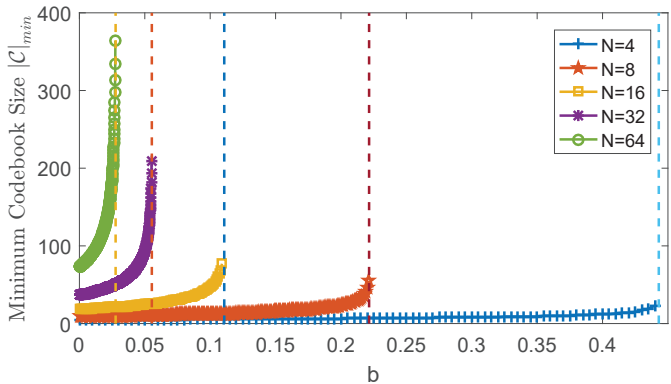}
\end{center}
\caption{Minimum codebook size $\left| \mathcal{C} \right|_{min}$ required to cover $\psi \in \left[ -1, 1 \right]$ as a function of fractional bandwidth $b$. The dashed lines indicate upper bounds. $g_t$ is set to be 3~dB smaller than $g_m$. $d=\lambda_c/2$.}
\label{fig:codebook-size-vs-b}
\end{figure}

Fig.~\ref{fig:codebook-design-without-wideband-effect} and \ref{fig:codebook-design-example} show examples of codebook designs for the minimum codebook size without and with considering beam squint, respectively, for $N=16$, $\psi_m=1$ and $d=\lambda_c/2$. $\psi_m=1$ denotes that the maximum targeted covered AoA/AoD of the codebook is $90^{\circ}$. If beam squint is considered, the minimum codebook size increases. The beams are denser for higher focus angles, because of the larger reduction of effective beamwidth for a larger beam focus angle as shown in (\ref{effective-beam-width}). In the two figures, the fractional bandwidth $b$ corresponds to $B=2.5$~GHz and $f_c=73$~GHz. The minimum codebook size in the examples for a $N=16$ ULA is increased by 15.8\% compared to that without considering beam squint, and it is increased by 54\% for a $N=32$ ULA with the same $b$.

Fig.~\ref{fig:codebook-size-vs-b} illustrates examples of the minimum codebook size $\left| \mathcal{C} \right|_{min}$ as a function of fractional bandwidth $b$ for $\psi_m=1$, i.e., the maximum targeted covered AoA/AoD of the codebook is $90^{\circ}$.
$b=0$ is equivalent to not considering beam squint.  For all $N$, as $b$ increases, $\left| \mathcal{C} \right|_{min}$ increases, and $b$ is upper bounded for fixed $N$, confirming Theorem~\ref{theorem:b-N-upper-bounded}. As $b$ approaches to its unreachable upper bound $\frac{\Delta\psi_{3\text{dB}}}{\psi_m}$, $\left| \mathcal{C} \right|_{min}$ increases significantly. If $b \to \frac{\Delta\psi_{3\text{dB}}}{\psi_m}$, the effective beamwidth of the beam with $\psi_0=\psi_m$, $\Delta\psi_{3\text{dB}} \left(\psi_0\right) \to 0$; therefore, $\left| \mathcal{C} \right|_{min}{ \to \infty }$. If $b \ge \frac{\Delta\psi_{3\text{dB}}}{\psi_m}$, no codebook exists.

\begin{figure}
\begin{center}
\includegraphics[width=88 mm]{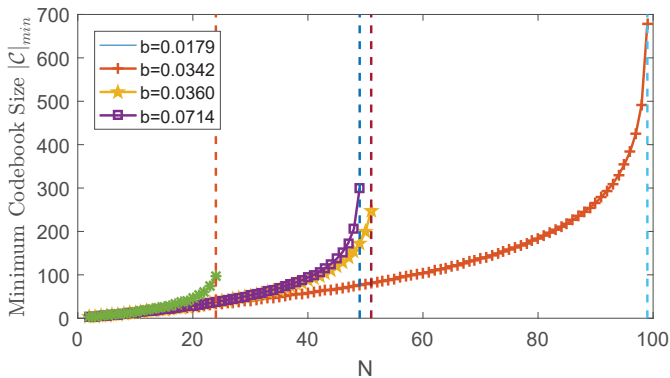}
\end{center}
\caption{Minimum codebook size $\left| \mathcal{C} \right|_{min}$ required to cover $\psi \in \left[ -1, 1 \right]$ as a function of the number of antennas $N$. The dashed lines indicate upper bounds. $g_t$ is set to be 3~dB smaller than $g_m$. $d=\lambda_c/2$. $b=0.0179$ corresponds to $B=0.5$ GHz and $f_c=28$ GHz; $b=0.0342$ corresponds to $B=2.5$ GHz and $f_c=73$ GHz; $b=0.0360$ corresponds to $B=2.16$ GHz and $f_c=60$ GHz; $b=0.0714$ corresponds to $B=2$ GHz and $f_c=28$ GHz.}
\label{fig:codebook-size-vs-N}
\end{figure}

Fig.~\ref{fig:codebook-size-vs-N} shows examples of the minimum codebook size $\left| \mathcal{C} \right|_{min}$ as a function of the number of antennas $N$ for $\psi_m=1$, i.e., the maximum targeted covered AoA/AoD of the codebook is $90^{\circ}$. $\left| \mathcal{C} \right|_{min}$ increases as $N$ increases. $N$ is upper bounded by $\floor[\big]{\frac{1.772}{\psi_m b}}$ as described in (\ref{N-upper-bounded}). When $N$ goes beyond the upper bound, no codebook exists. The maximum array gain only depends on $N$. Therefore, the maximum array gain $g_m=\sqrt{N}$ is also upper bounded.

\iftrue
\section{Conclusion}
We analyzed the effect of beam squint in analog beamforming with a phased array implemented via phase shifters as well as its influence on codebook design. Overall, the effect of beam squint increases if the AoA/AoD is away from the beam focus angle, and it tends to increase as AoA/AoD increases. We also developed an algorithm to compensate for the beam squint in codebook design. Beam squint increases the codebook size if the minimum array gain is required for all subcarriers in the wideband signal. The fractional bandwidth is upper bounded for fixed number of antennas; the number of antennas in the phased array is correspondingly upper bounded for fixed fractional bandwidth.


\fi


\balance


\bibliographystyle{IEEEtran}
\bibliography{IEEEabrv,mmc}
\end{document}